\documentclass[apj]{emulateapj}
\usepackage{multirow,amsmath,natbib,longtable,appendix}
\shorttitle{Compact massive spheroids}
\shortauthors{Graham et al.}

\begin{document}

\title{Hiding in plain sight: an abundance of compact massive spheroids in the local Universe}

\author{Alister W.\ Graham\altaffilmark{1}, Bililign T.\ Dullo, Giulia A.D.\ Savorgnan}
\affil{Centre for Astrophysics and Supercomputing, Swinburne University of
  Technology, Hawthorn, VIC, 3122, Australia.}
\altaffiltext{1}{Corresponding Author: AGraham@swin.edu.au}

\begin{abstract}
It has been widely remarked that compact, massive, elliptical-like galaxies
are abundant at high redshifts but exceedingly rare in the Universe today,
implying significant evolution such that their sizes at $z\sim 2\pm0.6$ have
increased by factors of 3 to 6 to become today's massive elliptical galaxies.
These claims have been based on studies which measured the half-light radii of
galaxies as though they are all single component systems.  Here we identify 21
spheroidal stellar systems within 90 Mpc that have half-light, major-axis 
radii $R_{\rm e} \lesssim 2$ kpc, stellar masses $0.7\times 10^{11} <
M_*/M_{\odot} < 1.4\times 10^{11}$, and S\'ersic indices typically around a
value of $n=2$ to 3. 
This abundance of compact, massive spheroids in our own
backyard --- with a number density of $6.9\times 10^{-6}$ Mpc$^{-3}$ (or
3.5$\times10^{-5}$ Mpc$^{-3}$ per unit dex in stellar mass) ---  and with
the same physical properties as the high-redshift galaxies, had been
over-looked because they are encased in stellar disks which usually result in
{\it galaxy} sizes notably larger than 2 kpc.  
Moreover, this number density is a lower limit because it has not come from 
a volume-limited sample. The actual density may be closer to $10^{-4}$,
although further work is required to confirm this. 
We therefore conclude that not
all massive `spheroids' have undergone dramatic structural and size evolution
since $z \sim 2\pm0.6$.  Given that the bulges of local early-type disk
galaxies are known to consist of predominantly old stars which existed at
$z\sim 2$, it seems likely that some of the observed high redshift spheroids
did not increase in size by building (3D) triaxial envelopes as commonly
advocated, and that the growth of (2D) disks has also been important 
over the past 9--11 billion years.  
\end{abstract}


\section{Introduction}

A little over a decade ago it was advocated that the mass-size relation for
massive galaxies ($M_* > 0.2\times 10^{11}\, h^{−2}_{70}\, M_{\odot}$) had
evolved little from $z\sim 2.5$ to today (Trujillo et al.\ 2004).  However
this view was quickly challenged by Daddi et al.\ (2005), who had detected
seven passively evolving galaxies at $z>1.4$, with stellar masses $\gtrsim
10^{11}\, h^{−2}_{71}\, M_{\odot}$ and morphologies typical of
elliptical/early-type galaxies.  They questioned the lack of evolution because
some of their galaxies had much smaller effective half light radii than local
elliptical galaxies of the same mass (see also Papovich et al.\ 2005).  They
suggested this difference was either a real sign of galaxy size evolution, or
artificial because of biasing active galactic nuclei light or morphological
$K$-corrections due to blue cores in the high-$z$ sample.  Addressing the
latter uncertainty, Kriek et al.\ (2006, 2008) found that nearly half of their
distant massive galaxies had old stellar populations and van Dokkum et
al.\ (2008) subsequently showed that half of those had sizes less than $\sim$2
kpc.  Following Daddi et al.\ (2005), Trujillo et al.\ (2006a) reported galaxy
sizes ranging from 1 to 5 kpc for ten massive ($\sim$5$\times 10^{11}\,
h^{−2}_{70}\, M_{\odot}$) galaxies at $1.2 < z < 1.7$.  From this Trujillo et
al.\ (2006a) concluded that {\it (i)} the sizes were at least a factor of 3 to
6 times lower than ``the local counterparts'' of similar mass, {\it (ii)} the
structural properties of these high-$z$ objects are therefore rapidly
changing, {\it (iii)} the data disagree with a scenario where the more massive
and passive galaxies are fully assembled by $1.2 < z < 1.7$ (i.e.\ a
monolithic scenario), and {\it (iv)} they suggested that a dry merger scenario
(no new star formation) was responsible for the subsequent evolution of these
galaxies (see also Trujillo et al.\ 2007; Toft et al.\ 2007; Zirm et
al.\ 2007; Buitrago et al.\ 2008; van Dokkum et al.\ 2008; Damjanov et
al.\ 2009).


While not denying the occurrence of dry mergers, nor claiming that early-type
galaxies had to have formed in a monolithic collapse (see Dekel \& Burkert
2014 and Zolotov et al.\ 2015 for an alternative), given the data under investigation we are questioning the 
universality of the above conclusions because they rest on the assumption as
to what the ``local counterparts'' to the compact, massive high-$z$ galaxies
actually are.  Popular opinion has been that these distant objects are very
rare today (e.g.\ Cimatti et al.\ 2008, and references therein), with Trujillo
et al.\ (2009) reporting a local number density of $10^{-7}$ Mpc$^{-3}$, and
Taylor et al.\ (2010) finding none and therefore concluding that the distant spheroidal
stellar systems have experienced considerable size evolution to form
today's massive elliptical galaxies.  However, there are many independent
lines of reasoning to think that this paradigm of early-type galaxy evolution
may not be correct. 

Key to much of this is that Trujillo et al.\ (2006a,b, 2007) used the sizes of
$z\sim 0.1$ {\it Sloan Digital Sky Survey} ({\it SDSS}, York et al.\ 2000)
early-type, i.e.\ elliptical and lenticular, {\it galaxies} (with S\'ersic
index $n>2.5$), to claim that there has been significant size, and
surface-mass density, evolution of the distant spheroids (see also Taylor et
al.\ 2010; Newman et al.\ 2012; 
McLure et al.\ 2013; Damjanov et al.\ 2014; Fang et al.\ 2015).  In
addition to early-type galaxies, there is however another type of spheroidal
stellar system in the local universe, namely the bulges of disk galaxies,
including the lenticular galaxies. Furthermore, there has been a slowly
growing realization over the past three decades that most\footnote{The
  exceptions are brightest cluster galaxies and luminous ($M_B \lesssim
  -20.5\pm0.75$ mag) elliptical galaxies built via major dry merger events,
  and very faint dwarf elliptical galaxies.} early-type galaxies infact
consist of a bulge and a rotating stellar disk (e.g.\ Carter 1987; Capaccioli
1987, 1990; Nieto et al.\ 1988; D'Onofrio et al.\ 1995; Graham et al.\ 1998;
Emsellem et al.\ 2011; Scott et al.\ 2014), arguing against the treatment of
lenticular galaxies as single component systems.  While many of these galaxies
do not have massive bulges, and some contain pseudobulges built out of disk
material (Kormendy \& Kennicutt 2004; see also Graham 2014 for cautionary
remarks regarding their identification), our interest lies with some of the
more massive lenticular and spiral galaxies that do have massive bulges. The
most massive local bulges ($M_* > 2\times 10^{11} M_{\odot}$) have sizes in
excess of 2.5 kpc are therefore not considered ``compact''; as such they have
been excluded from this study.

Here we effectively explore what evolutionary conclusion might have been drawn
a decade ago for the compact massive spheroids at $z\approx 2$ if the
size-mass comparison had of been performed with the massive bulges of
present-day early-type disk galaxies rather than with the combined bulge+disk
system.  In so doing we can answer the question, Have the high-$z$ spheroidal
stellar systems truly evolved to near extinction, or are they perhaps hiding
in plain sight around us today?  While we do not deny that there has been {\it
  galaxy} size evolution, we are suggesting a fundamentally different
formation model which presents a dramatic shift in our way of thinking about
how the compact massive objects at $z = 2.0\pm0.6$ have actually evolved in
our Universe.  It is such that they have not expanded 3--6 times in size, nor
accreted, through dry minor mergers, a three-dimesnional (3D) envelope of a
similarly large dimension, to become today's massive elliptical galaxies
(e.g.\ Hopkins et al.\ 2009; Carrasco et al.\ 2010; Cimatti et al.\ 2012; Fan
et al.\ 2013b, De et al.\ 2014).  Instead, we advocate the view presented by Graham (2013) that
they have grown two-dimensional (2D) planar disks and are thus effectively
hidden today until bulge-to-disk decompositions are performed.  It turns out
that there are actually many reasons to favour this scenario.

There is no widely accepted solution as to how the distant objects could have
grown into much larger `spheroids' by today.  For instance, major dry mergers
can not account for the size evolution; they move galaxies {\it along} the
mass-size relation rather than off it (e.g.\ Ciotti \& van Albada 2001;
Boylan-Kolchin et al.\ 2006; see also Bundy et al.\ 2009; Nipoti et al.\ 2009;
Nair et al.\ 2011).  In addition, based on observations of galaxy pairs, there
have been insufficient major (i.e.\ near equal mass ratio) merger events to
explain the removal of every distant, compact massive galaxy (e.g.\ Man et
al.\ 2015; see also Shih \& Stockton 2011).  Similarly, there are not
enough satellites observed around massive galaxies for minor mergers
(e.g.\ Khochfar \& Burkert 2006; Maller et al.\ 2006; Naab et al.\ 2009;
Hopkins et al.\ 2009; McLure et al.\ 2013)
%
%
to be the sole explanation (Trujillo 2013, and references therein), 
%
%
and even in $\Lambda$CDM simulations there are not enough satellites to have
transformed all of the distant spheroids.

It is worth noting that `normal-sized' (i.e.\ not compact), massive
elliptical-like galaxies {\it are} observed at these high-redshifts,
co-existing with the `compact', massive galaxies (e.g.\ Trujillo et
al.\ 2006b; Mancini et al.\ 2010; Newman et al.\ 2010; Bruce et al.\ 2012;
Ferreras et al.\ 2012; Fan et al.\ 2013a).  Furthermore, new galaxies are
known to appear on the ``red sequence'' (e.g.\ Bell et al.\ 2004; Faber et
al.\ 2007; Schawinski et al.\ 2007, 2014; van Dokkum et al.\ 2008; Martig et
al.\ 2009; van der Wel et al.\ 2009; L\'opez-Sanjuan
et al.\ 2012; Carollo et al.\ 2013; Cassata et al.\ 2013; Keating et
al.\ 2015; Belli et al.\ 2015).  For example, Carollo et al.\ (2015) detail how
past quenching can transform star-forming galaxies and shift them to the
early-type sequence. 
Known as `progenitor bias', studies which attempt to measure the
evolution of the massive quiescent galaxies (e.g.\ Trujillo et al.\ 2006b; Ryan et al.\ 2012;
Chang et al.\ 2013; van der Wel et al.\ 2014) can suffer from sample confusion
because of the emerging population onto the red sequence; such studies are
effectively sampling different objects as a function of redshift.

In mid-2011, Graham (2013, see also Driver et al.\ 2013) suggested that some
of the distant objects are likely to be the bulges of today's massive
early-type disk galaxies.  Indeed, Graham (2013) showed that the distant compact massive
objects and massive local bulges occupy the same region in the size-mass and
size-density diagram.  Dullo \& Graham (2013) additionally showed that they
possess the same radial concentration of light, as traced by the S\'ersic
index (S\'ersic 1968, see Graham \& Driver 2005 for a modern review in
English).  Moreover, massive bulges today are old.  Their mass (not to be
confused by their luminosity-weighted age) is dominated by old stars
(MacArthur et al.\ 2009) and therefore such luminous bulges {\it should} be
visible in images of the early universe.  That is, some fraction of the
compact massive objects reported at $z\sim 2\pm0.6$ are expected to be today's
compact massive bulges.  By calculating the number density of compact massive
bulges at $z\sim 0$, done for the first time in this work, we are able to
address whether this fraction is low or instead if it might equal 1 and thus 
fully account for the fate of the distant compact massive objects.

Rather than some great mystery of unexplained galaxy growth, we are exploring
the relatively mundane alternative idea that Daddi et al.\ (2005) and Trujillo
et al.\ (2006a,b) simply observed the bulges of today's bright early-type
galaxies at high-$z$ but did not know it.  To address this, we have performed
a preliminary search for nearby ($\lesssim 100$ Mpc), early-type disk galaxies
with bulges having $R_{\rm e} \lesssim 2$ kpc and $M_* \approx 10^{11}
M_{\odot}$.  This was done by checking on the galaxies presented in a handful
of published papers listed in Table~\ref{Tab1}.  This proved to be 
sufficient to illustrate that compact, massive spheroids exist in substantial
numbers around us today.  Section~2 of this paper presents the galaxies found,
and the structural properties of their spheroidal component are also provided
there.  Section~3 provides a general discussion of these findings and the
growth of disks in our Universe.  Section~4 finishes by re-iterating our main
conclusions.


\section{Data Sample}\label{Sec_data}

Five survey papers were consulted to check for the existence of compact
massive bulges in the nearby Universe.  Our criteria was that the stellar mass
be greater than $0.7\times10^{11} M_{\odot}$ and the major-axis half-light
radius be less than 2.5 kpc\footnote{It should be noted that studies of
  the high`$z$ galaxies typically report the smaller, circularized half-light radii rather
  than the major-axis half-light radii.  Our approach is thus conservative.}. 
The first paper we examined was Dullo \& Graham (2013)
which has already broached this topic, and the second was Savorgnan et
al.\ (2015, in preparation) which has carefully modeled the different
structural components for 66 galaxies from a sample of 75 (Graham \& Scott 2013; Scott
et al.\ 2013) having directly measured black hole masses.  Building on this, we
checked two samples dominated by lenticular galaxies.  First we inspected the
sample of 175 early-type disk galaxies modeled by Laurikainen et al.\ (2010),
and then the ATLAS$^{\rm 3D}$ sample (Cappellari et al.\ 2011) of 260 nearby
(D $<$ 42 Mpc), predominantly northern hemisphere, early-type galaxies which
have had the stellar bulge-to-total flux ratios derived by Krajnovi\'c et
al.\ (2013).  The fifth catalog paper that we inspected was the spiral
dominated compendium of Graham \& Worley (2008), which resulted in the
identification of targets in four additional papers (see Table~\ref{Tab1}).
Having checked these papers, our Table~\ref{Tab1} includes 20 known 
bulges along with the known
compact massive {\it galaxy} NGC 1277 (van den Bosch et al.\ 2012),
plus one additional {\it galaxy}, as opposed to a bulge, which also meets the above
size and mass criteira.  All but three of the 22 systems have $R_{\rm e} < 2$
kpc.  All but 4 have $M_* \ge 0.9\times 10^{11} M_{\odot}$. 

The major-axis, half-light radii of the spheroids are presented in
Table~\ref{Tab1}, based on the galaxy distances which are also listed there.
The circularized half-light radii would be even smaller, by a factor of
$\sqrt{b/a}$ where $b/a$ is the minor-to-major axis ratio of the bulge. This
information was not readily available for many of our bulges, and so we have
used the more conservative, larger radii.  The published S\'ersic indices have
additionally been collated and given in Table~\ref{Tab1} for ease of
reference.  Our preference has been for near-infrared data sets because the
magnitudes are more reliable: they are less affected by dust and possible low
level star formation.  The absolute magnitudes were then corrected for
redshift dimming ($5\log[1+z]$), foreground Galactic extinction (Schlafly \&
Finkbeiner 2011), and $K$-corrected ($1.5z$, Poggianti 1997) which resulted in
insignificant changes of typically 0.01--0.02 mag.

In the case of NGC~5493 (Krajnovi\'c et al.\ 2013), the whole galaxy is rather
compact ($R_{\rm e} = 2.46$ kpc) --- although this is the largest system in our
sample --- and massive ($M_* = 10^{11} M_{\odot}$)\footnote{Krajnovi\'c et
al.\ (2013) reported that their bulge/disk decomposition for NGC~5493 differed 
markedly from Laurikainen et al.\ (2010). Given this uncertainty, we use the 
whole galaxy rather than a result from a decomposition.}.  This galaxy is not unique in
the nearby universe. The intermediate-scale disk in NGC~1332, which does not
dominate at large radii like large-scale disks do, is such that the size of
this {\it galaxy} is only slightly larger than 2 kpc (Savorgnan et al.\ 2015, in preparation).
However here we include the bulge rather than the galaxy parameters for
NGC~1332.  A third and possible fourth example is NGC~1277 ($M_* =
1.2\times10^{11} M_{\odot}$) and NGC~5845 (excluded here because $M_* =
0.5\times10^{11} M_{\odot}$), as already pointed out by van den Bosch et
al.\ (2012) and Jiang et al.\ (2012), respectively.  However, if the disks of
these galaxies were to continue to grow, until they resembled the more
stereotypical S0 galaxies today, with $B/D$ flux ratios of 1/3 and $R_{\rm
  e,bulge}/h_{\rm disk} \approx 0.2$ ($R_{\rm e,bulge}/R_{\rm e,disk} \approx
0.12$), then their galaxy sizes would likely exceed 2 kpc while the spheroidal
component remained compact.
 
For reference, it was observed that the bulges with stellar masses $M_*
\gtrsim 2\times 10^{11} M_{\odot}$ tend to have major-axis, half-light radii
$R_{\rm e} > 2.5$ kpc, and as such were not included here.

\begin{deluxetable*}{lccccccccccc}
\footnotesize
\tablecaption{Galaxy Sample. \label{Tab1}}
\tablewidth{0pt}
\tablehead{
Gal.        & Type           & Inc.  &  z     &  Dist     & $n_{\rm bulge}$ & $R_{\rm e,bulge}$ & $M_{\rm bulge}$ & $M^{\rm corr1}_{\rm bulge}$ & $M^{\rm corr2}_{\rm bulge}$ & $M/L$ & Mass$^{\rm corr}_{\rm *, {\rm bulge}}$  \\ 
Id.         &                & [deg] &        & [Mpc]     &              & [kpc]          &   [mag]      &      [mag]             &   [mag]               &       & [$10^{11}\, M_{\odot}$]               \\
(1)         & (2)            & (3)   & (4)    & (5)       & (6)          & (7)            &   (8)        &      (9)               &      (10)             & (11)  &   (12)                           \\
}
\startdata
\multicolumn{12}{c}{Lenticular galaxies} \\
\hline 
\multicolumn{12}{c}{Laurikainen et al.\ (2010, their table~1) $K_s$-band bulge magnitude} \\
ESO 337-G10  & SA(s)0$^-$     & 37.6 & 0.0193 & 79.0      & 1.7          &  1.52          & $-24.33$      &  $-24.37$             &       ...             &  0.8  &  0.9   \\
NGC~484      & SA0$^-$        & 45.1 & 0.0170 & 68.4      & 2.8          &  1.69          & $-24.63$      &  $-24.65$             &       ...             &  0.8  &  1.2   \\
NGC~1161     & SA(l)0$^0$     & 33.6 & 0.0065 & 28.4      & 2.5          &  1.77          & $-24.20$      &  $-24.27$             &       ...             &  0.8  &  0.8   \\
NGC~3665     & SA0$^0$(s)     & 39.1 & 0.0069 & 32.4      & 2.7          &  1.98          & $-24.36$      &  $-24.37$             &       ...             &  0.8  &  0.9   \\
NGC~5266     & SA0$^-$        & 54.2 & 0.0100 & 40.8      & 2.8          &  1.84          & $-24.79$      &  $-24.82$             &       ...             &  0.8  &  1.4   \\
NGC~5419     & SA(nl)0$^-$    & 43.9 & 0.0138 & 54.8      & 1.4          &  1.74          & $-24.60$      &  $-24.63$             &       ...             &  0.8  &  1.2   \\
NGC~7796     & S0             & 27.9 & 0.0112 & 48.5 [Ton]& 2.2          &  1.72          & $-24.47$      &  $-24.48$             &       ...             &  0.8  &  1.0   \\
\multicolumn{12}{c}{ATLAS3D: $r'$-band $R_{\rm e}$ and $n$. $B/T \times K_s$-band galaxy luminosity (Cappellari et al.\ 2011, their table~3; Krajnovi\'c et al.\ 2013)} \\
NGC~4649$^{\dag}$ & S0       & 39.6 & 0.0037 & 17.3      & 1.8          & 1.42$\pm$0.10  & $-24.22$      &  $-24.22$             &       ...             &  0.8  &  0.80  \\  
NGC~3640     & S0            & 34.9 & 0.0043 & 26.3      & 2.1          & 2.12$\pm$1.11  & $-24.05$      &  $-24.05$             &       ...             &  0.8  &  0.7   \\
NGC~5493$^*$ & S0            & 65.2 & 0.0089 & 38.8      & 5.8$^*$      & 2.46$\pm0.11^*$ & $-24.49^*$   &  $-24.49^*$           &       ...             &  0.8  &  1.0$^*$ \\  
\multicolumn{12}{c}{Savorgnan et al.\ (2015, in preparation) 3.6 $\mu$m bulge magnitude} \\
NGC~1316     & SAB(s)0$^0$    & 45   & 0.0059 & 20.0 [B09]& 2.0$^{**}$   &  1.54$^{**}$   & $-25.05^{**}$  &  $-25.05^{**}$         &       ...             &  0.6  &  1.2    \\
NGC~1332     & S0$^-$(s)      & 73   & 0.0051 & 22.3 [Ton]& 5.3$^{**}$   &  1.95$^{**}$   & $-24.89^{**}$  &  $-24.89^{**}$         &       ...             &  0.6  &  1.1    \\
\multicolumn{12}{c}{Dullo \& Graham (2013)  $V$-band $R_{\rm e}$ and $n$. $B/T_{\rm Obs} \times$ {\sc 2MASS} $K_s$-band galaxy luminosity} \\
NGC~507      & SA(r)0$^0$     &  0.0 & 0.0165 & 66.2 [NED] & 2.2         &  1.71          & $-24.57$      &  $-24.60$             &       ...             &  0.8  &   1.1  \\  
\multicolumn{12}{c}{van den Bosch et al.\ (2012)} \\
NGC~1277$^*$ & S0$^+$pec      & 75   & 0.0169 & 73.0      & 2.2$^*$      &  1.2$^*$       &  ...          &  ...                  &       ...             &       &  1.2$^*$ \\ 
\hline
\multicolumn{12}{c}{Spiral galaxies} \\
\hline 
\multicolumn{12}{c}{Graham (2001, 2003) $K$-band bulge magnitude} \\
NGC~266      & SB(rs)ab       & 27.1 & 0.0155 & 62.8 [NED]& 1.4         & 1.25           & $-24.14$       &  $-24.17$             &      $-24.28$         &  0.8  &  0.9    \\
NGC~2599     & SAa            & 29.5 & 0.0158 & 68.4 [NED]& 2.1         & 1.71           & $-24.45$       &  $-24.47$             &      $-24.58$         &  0.8  &  1.1    \\
NGC~3884     & SA(r)0/a       & 40.5 & 0.0234 &103.6 [NED]& 2.3         & 1.64           & $-24.41$       &  $-24.43$             &      $-24.55$         &  0.8  &  1.1    \\
NGC~7490     & Sbc            & 24.5 & 0.0207 & 84.9 [NED]& 2.6         & 1.53           & $-23.99$       &  $-24.03$             &      $-24.14$         &  0.8  &  0.7    \\
\multicolumn{12}{c}{Seigar \& James (1998) $K$-band bulge magnitude} \\
UGC~2862     & (R')SAB(s)a    & 44.4 & 0.0224 & 89.3 [NED]& 1.9         & 1.15           & $-24.22$       &  $-24.34$             &    $-24.48$           &  0.8  &  1.0    \\
\multicolumn{12}{c}{M\"ollenhoff \& Heidt (2001) $K$-band bulge magnitude} \\
NGC~3147     & SA(rs)bc       & 34.9 & 0.0093 & 43.0 [NED]& 3.6         & 1.39           & $-24.41$       &  $-24.42$             &    $-24.54$           &  0.8  &  1.1    \\
\multicolumn{12}{c}{Balcells et al.\ (2007, their table~2) $K$-band bulge magnitude} \\
NGC~6504     & Sab            & 78.5 & 0.0160 & 72.1 [NED]& 2.7         &  1.20          & $-24.11$       &  $-24.13$             &    $-24.67$           &  0.8  &  1.2    \\
\multicolumn{12}{c}{Laurikainen et al.\ (2010) $K_s$-band bulge magnitude} \\ 
NGC~6646     &(R')SAB(rs)a    & 23.4 & 0.0192 &  81.8     & 2.2         &  2.16          & $-24.16$       &  $-24.19$             &     $-24.30$          &  0.8  &   0.9   \\
\enddata
\tablecomments{
Column 1: Galaxy identification. 
* The galaxy rather than bulge parameters are presented. 
\dag NGC~4649 (M60) is a particularly difficult galaxy to decompose and as
such its bulge parameters may not be reliable. 
Column 2: Morphological Type. 
Column 3: Disk inclination such that 90 degrees corresponds to an edge-on disk. 
Column 4: Redshift taken from the NASA/IPAC Extragalactic Database (NED). 
Column 5: Distance from corresponding paper unless otherwise specified: 
NED = (Virgo + GA + Shapley) distance from NED using $H_0=73$ km s$^{-1}$ Mpc$^{-1}$); 
B09 = Blakeslee et al.\ (2009); 
Ton = Tonry et al.\ (2001) and corrected according to Blakeslee et al.\ (2002). 
Column 6: S\'ersic index. 
Column 7: Major-axis, effective half-light radius of the galaxy's bulge component. 
Column 8: Near-infrared magnitude. Notes: 
* = galaxy rather than bulge parameters; 
** 3.6 $\mu$m magnitude rather than $K$- or $K_s$-band. 
Column 9:  Magnitude corrected for Galactic dust extinction (Schlafly \& Finkbeiner 2011), $(1+z)^2$ cosmological dimming, and $K$-corrected using $+1.5z$ (Poggianti 1997). 
Column 10: Magnitude additionally corrected for internal dust using the correction from Driver et al.\ (2008). 
Column 11: Stellar mass-to-light ratio (assumes a 12 Gyr old population of
solar metallicity and a Chabrier (2003) initial mass function: Baldry et al.\ 2008,
their Figure~A1). 
Column 12: Stellar mass derived using column~9 for the S0s and column~10 for
the spirals, together with column~11.
}
\end{deluxetable*}

We elected not to apply any internal dust correction to the lenticular
galaxies, but only to the spiral galaxies.  The most massive lenticular
galaxies are not representative of the typical dusty disk galaxy from which the 
dust corrections stem, and they likely contain little wide-spread dust. 
The following $K$-band dust correction from Driver et al.\ (2008) was applied
to the bulges of the spiral galaxies. 
\begin{equation} 
M^{\rm corr}_{\rm bulge} = M_{\rm bulge} -0.11 - 0.79[1-\cos(i)]^{2.77}, 
\end{equation}
where $i$ is the inclination of the disk such that $i=90$ deg corresponds to
an edge-on orientation.  For the Dullo \& Graham (2013) sample of five
lenticular galaxies, the stellar masses derived from the $V$-band magnitudes
(RC3, de Vaucouleurs et al.\ 1991) agree with those derived from the $K_s$-band magnitudes when no dust
correction is applied and $M/L_{K_s} = 0.8$ and $M/L_V =2.5$ is used.  
%
%
While the Sombrero lenticular galaxy contains wide-spread dust in its disk, as
do other lenticular galaxies (e.g.\ Temi et al.\ 2007), it may be that ion
sputtering (e.g.\ Draine 2003) from a hot X-ray gas halo has destroyed the
dust in the massive lenticular galaxies, although we have not checked for such
X-ray halos in our sample.  However, if we are mistaken and dust is present,
it will mean that our estimated stellar masses should be increased.  As shown
in Driver et al.\ (2008), the average correction due to dust in the thousands
of disk galaxies modeled as a part of the Millennium Galaxy Catalog (Liske et
al.\ 2003; Allen et al.\ 2006) is $<$0.14 mag if the inclination is $<$45$^o$,
and 0.18, 0.28 and 0.45 mag if the inclination is 55, 65 and 75 degrees,
respectively.

Some of the near-infrared magnitudes in Table~\ref{Tab1} were derived from Two
Micron All-Sky Survey (2MASS) images (Skrutskie et al.\ 2006).  As noted by
Schombert \& Smith (2012), the 2MASS total magnitudes are known to not capture
all of a galaxy's light; they miss flux at large radii.  However, for most S0s
this appears to be contained to, on average, a tenth of a magnitude (Scott et
al.\ 2013, their Figure~2), no doubt due to the rapid decline of the S0
galaxy's outer exponential light profile.  For S0s with intermediate
(i.e.\ not large) scale disks that do not dominate their galaxy's light at
large radii, and if the S\'ersic index of the bulge is large, then several
tenths of a mag may be missed, as with the elliptical galaxies.  The (Galactic
extinction)-corrected $B-K_s$ colors (via NED but from the RC3, 2MASS and
Schlafly \& Finkbeiner 2011) range from about 3.65 to 4.05.  Given the
expected color range of 3.85 to 4.03 for a 6 to 12.5 Gyr old population in
model S0 galaxies (Buzzoni 2005), {\it perhaps} a couple of tenths of mag are
missing from some 2MASS magnitudes.  If one was to correct for this, it too
would act to make the bulge masses bigger than reported here.


A Hubble constant of $H_0 = 73$ km s$^{-1}$ Mpc$^{-1}$ (e.g.\ WMAP 3-year;
Riess et al.\ 2011) was used. However if it is smaller, for example 67.3 km
s$^{-1}$ Mpc$^{-1}$ (Planck Collaboration 2014), then the absolute magnitudes
will brighten by 0.18 mag and the stellar masses will increase by 18\%.  Based
on SNe~Ia data, Rigault et al.\ (2015) have also downwardly revised $H_0$ to
$70.6\pm 2.6\ \mathrm{km\ s^{-1}\ Mpc^{-1}}$.  A mass increase of 18\% would
result in 18 of the 22 systems in Table~\ref{Tab1} having stellar masses 
$\ge 10^{11} M_{\odot}$.

\subsection{Example profile: NGC~5419}

While information about the surface brightness profile decompositions can be
found in the papers mentioned in Table~\ref{Tab1}, we felt that it may be
instructive to include an example of how the outer exponential profile can
increase the galaxy size over the bulge size.  We have somewhat randomly
chosen NGC~5419 from Laurikainen et al.\ (2010). 
This was chosen because a) the Laurikainen et al.\ paper simply contains the
greatest number of compact massive spheroids in our Table~\ref{Tab1}, and b) this
particular galaxy was modelled by Dullo \& Graham (2014) as an elliptical 
galaxy while they noted that a disk might be present.  
Sandage \& Tammann (1981) did not support the original elliptical
galaxy classification in the Second Reference Catalogue of de Vaucouleurs et
al.\ (1976), but instead considered NGC~5419 to be a lenticular galaxy, as did
Laurikainen et al.\ (2010) based on their near-infrared $K$-band analysis.

\begin{figure}
  \includegraphics[scale=0.6,angle=0]{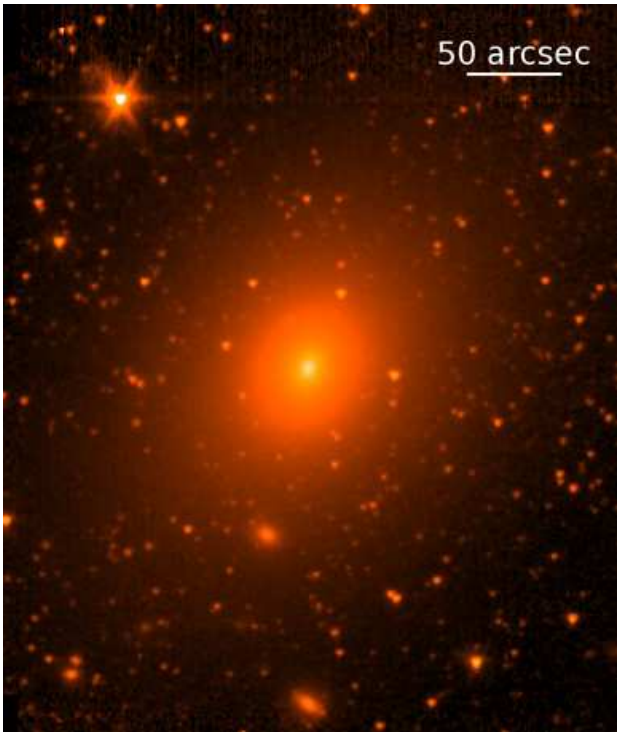}
  \caption{{\it Spitzer Space Telescope} 3.6$\mu$m image of NGC~5419. 
}
\label{Fig1}
\end{figure}
\begin{figure}
  \includegraphics[scale=0.47,angle=0]{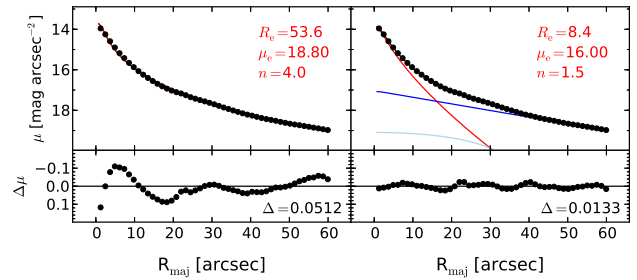}
  \caption{Left panel: Single S\'ersic fit to the 3.6$\mu$m, major-axis light
    profile of NGC~5419.  Right panel: As in Laurikainen et al.\ (2010), a
    S\'ersic bulge (red), plus exponential function (dark blue), plus Ferrers function
    for a lens (cyan) has been fit.  The root mean square of the
    deviations of the surface brightness data, $\mu$, about the model
    is given by $\Delta$.  }
\label{Fig2}
\end{figure}

Therefore, we re-investigate this galaxy using a {\it Spitzer Space Telescope}
3.6$\mu$m image (see Figure~\ref{Fig1}), which was reduced following the
procedures described in Savorgnan et al.\ (2015, in prep.).  The light profile
was extracted by allowing the ellipticity and position angle to vary about a
fixed center using the {\sc IRAF} task {\sc Ellipse} (Jedrzejewski et
al.\ 1987) and is shown in Figure~\ref{Fig2}.




Our modelling of the 3.6 $\mu$m light profile within 60$\arcsec$ matches the
solution to the $K$-band data in Laurikainen et al.\ (2010).  We also find
that a single S\'ersic function is inadequate; the curvature of the light
profile, and the residual light profile, reveal the presence of additional
components.  Figure~\ref{Fig2} shows how the addition of an outer exponential,
and a very faint lens (fit with a Ferrers function), provides a better fit.
These were the components employed by Laurikainen et al.\ (2010).  From this,
one can see that the half-light radius of the galaxy ($R_{\rm e,gal}=53\arcsec
.6$) is much greater than the half light radius of the inner spheroid ($R_{\rm
  e,sph}=8\arcsec .4$).  Laurikainen et al.\ (2010) reported that their
bulge+lens+disk model had a bulge S\'ersic index $n=1.4$ and half light radius
$R_{\rm e}=6\arcsec.5$.  They reported a disk scale length $h=32\arcsec.3$,
while our exponential model has $h=33\arcsec.1$.

One may wonder if there is also a contribution from an extended envelope
around this galaxy due its location at the centre of the poor cluster Abell
S753.  Indeed, Sandage \& Bedke (1994) referred to this galaxy as having an
extended outer envelope, and Seigar et al.\ (2007, see also Pierini et
al.\ 2008) established that intracluster light tends to have an exponential
profile, i.e.\ the same radial decline as seen in disks.  However, Bettoni et
al.\ (2001) have reported a rotational velocity reaching 90 km s$^{-1}$ by the
inner 5$\arcsec$, betraying the presence of at least an inner disk.  In an
extreme scenario, one may speculate that this galaxy has an intermediate-scale
disk (the `lens' component in Figure~\ref{Fig2}, which shows up in both the
ellipticity and position angle profiles) plus intracluster stars that produce
the near-exponential light profile at larger radii.  More extended kinematics
would be helpful in discriminating between these options.

The cluster-centric location of NGC~5419 is interesting, and it further
prompted us to pursue the suggestion by L.Cortese (2015, priv.\ comm.) to check
on the environment (e.g.\ field, group, cluster) of our sample.
Table~\ref{Tab2} shows this information, along with the $R_e/h$ size and $B/T$
flux ratios obtained by the papers given in Table~\ref{Tab1}.  While a few of
the galaxies are the brightest of their small galaxy group, NGC~1316 (the
interacting galaxy Fornax~A) is the only other member of a substantial-sized
group.  This rules out the idea that the compact massive spheroids might be
encased in an exponential-like 3D envelope of intracluster light rather than a
2D disk.

\begin{deluxetable}{lcccl}
\footnotesize
\tablecaption{Galaxy properties. \label{Tab2}}
\tablewidth{0pt}
\tablehead{
Gal.\ Id.\   & $\sigma$ & $R_{\rm e}/h$  &  $B/T$   &  Environment \\
}
\startdata
\multicolumn{5}{c}{Lenticular galaxies} \\
\hline 
\multicolumn{5}{c}{Laurikainen et al.\ (2010)} \\
ESO 337-G10  & ... &   0.25         &    0.41     &  ?    \\
NGC~484      & 233 &   0.29         &    0.62     &  Group \\
NGC~1161     & 295 &   0.40         &    0.65     &  ?     \\
NGC~3665     & 215 &   0.43         &    0.58     &  BGG, NGC~3665 Group ($N=11$) \\
NGC~5266     & 201 &   0.23         &    0.45     &  Group \\
NGC~5419     & 351 &   0.20         &    0.24     &  BCG, poor cluster Abell S753 \\
NGC~7796     & 259 &   0.27         &    0.48     &  Field   \\  
\multicolumn{5}{c}{Cappellari et al.\ (2011); Krajnovi\'c et al.\ (2013)} \\
NGC~4649$^{\dag}$ & 329 &  0.45     &    0.32    &  Virgo cluster ($N=355$) \\
NGC~3640     & 192 &     0.56       &    0.60    &  BGG, NGC~3640 Group ($N=12$) \\
NGC~5493$^*$ & 204 &     ...        &    ...     &  BGG, NGC~5493 Group ($N=2$) \\
\multicolumn{5}{c}{Savorgnan et al.\ (2015, in preparation)} \\
NGC~1316     & 225 &    0.08        &    0.23    &  BGG, NGC~1316 Group ($N=111$)  \\
NGC~1332     & 321 &    0.52        &   0.95     &  BGG, NGC~1332 Group ($N=22$) \\
\multicolumn{5}{c}{Dullo \& Graham (2013)} \\
NGC~507      & 295 &    0.19        &   0.32     &  BGG, poor NGC~507 Group  \\
\multicolumn{5}{c}{van den Bosch et al.\ (2012)} \\
NGC~1277$^*$ & 333 &     ...        &    ...     &  Group  \\
\hline
\multicolumn{5}{c}{Spiral galaxies} \\
\hline 
\multicolumn{5}{c}{Graham \& Worley (2008)} \\
NGC~266      & 230 &    0.19        &   0.24     &  Field?    \\
NGC~2599     & ... &    0.24        &   0.51     &  Field?    \\
NGC~3884     & 209 &    0.20        &   0.22     &  Abell~1367 cluster (N=90) \\      
NGC~7490     & ... &    0.23        &   0.27     &  Field?    \\    
UGC~2862     & ... &    0.11        &   0.55     &  Field?    \\ 
NGC~3147     & 225 &   0.33         &   0.23     &  BGG, NGC~3147 Group ($N=5$) \\
NGC~6504     & 185 &   0.23         &   0.42     &  Field?   \\
\multicolumn{5}{c}{Laurikainen et al.\ (2010)} \\ 
NGC~6646     & ... &   0.28         &   0.26     &  Field?    \\
\enddata
\tablecomments{
Column~1: Galaxy identification.
* The galaxy rather than bulge parameters were used. 
\dag Bulge/disk parameters may not be reliable.
\ddag NGC~1332 has had its disk fit with a S\'ersic model having $n=0.5$, for
which the scale length $h=R_{\rm e,disk}/0.82$ (rather than $R_{\rm
  e,disk}/1.68$ as is the case when $n=1$). 
Column~2: Velocity dispersion (km s$^{-1}$) from HyperLeda (Makarov et
al.\ 2014), except for NGC~1277 (van den Bosh et al.\ 2012). 
Columns~3 and 4: Bulge-to-disk size ratio and bulge-to-total flux ratio,
respectively.
For the lenticular galaxies (and NGC~6646), values were taken from the respective papers.
For the spiral galaxies, the dust-corrected values were derived in Graham \& Worley (2008).
Column~5: Environment of the galaxy. 
BCG = Brightest Cluster Galaxy.
BGG = Brightest Group Galaxy.
$N$ represents the number of galaxies with known radial velocities in the group (Makarov \&
Karachentsev 2011; Tully et al.\ 2013). 
Field? = No group known to authors, and the galaxy appears to be in the field from looking at
Digitized Sky Survey images available via NED.
}
\end{deluxetable}

\section{Results}

\subsection{The mass-size and size-concentration diagrams} 

\begin{figure*}
  \includegraphics[scale=0.99,angle=-90]{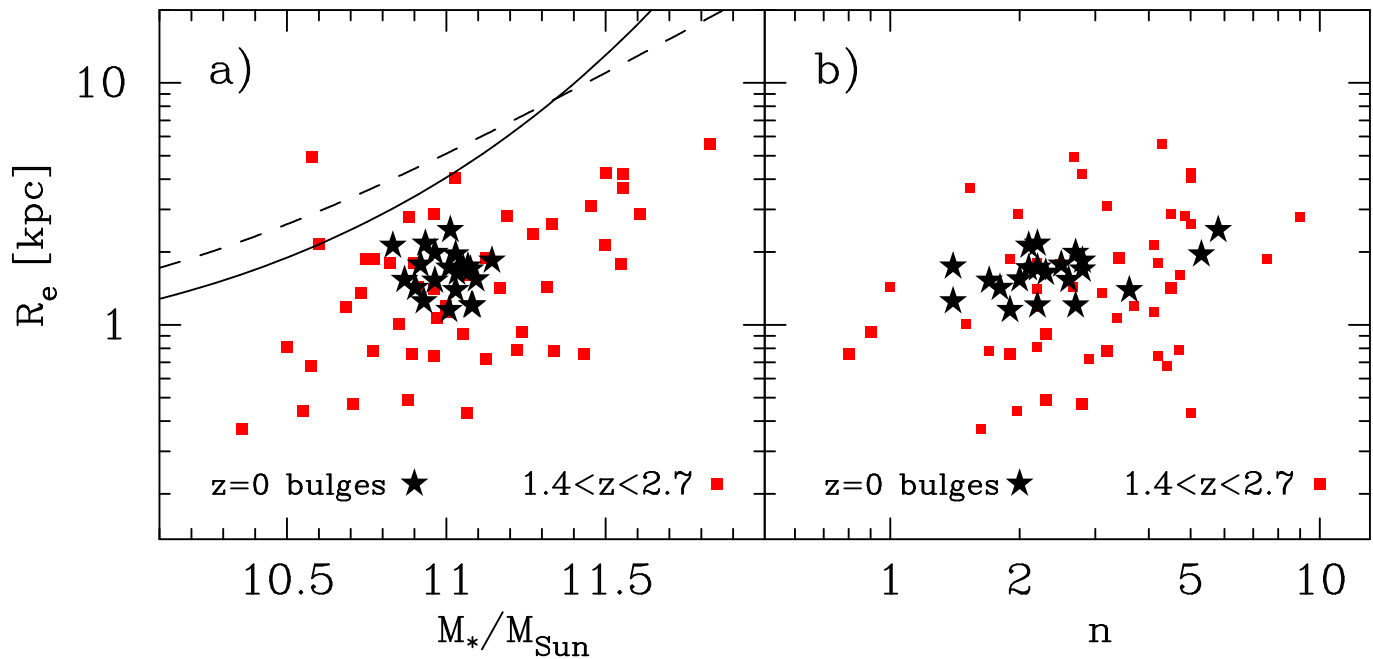}
  \caption{Panel a) Stellar mass--size diagram, where the sizes of the 
local bulges are represented here by the 
major-axis effective half light radius $R_{\rm e}$.  The curves show 
the parameterized fit to local ``elliptical'' galaxy data from 
Graham et al.\ (2006, solid line) and 
Lange et al.\ (2015, specifically their Figure~A9j and Eq.3, and their final $K$-band entry in
Table~3, dashed line).  
The $z=0$ bulge data is from Table~\ref{Tab1} while the $1.4< z < 2.7$ data is
from Damjanov et al.\ (2011). Local bulges with $M_* \gtrsim 2\times10^{11}
M_{\odot}$ have $R_{\rm e} \gtrsim 2.5$ kpc and were thus not included here
because they are not considered to be compact galaxies.
Panel b) 
Size-concentration diagram, where the concentration is quantified
by the S\'ersic index $n$.  
%
%
}
\label{Fig3}
\end{figure*}

In Figure~\ref{Fig3}a we do not compare our data with the size-mass relation
for early-type galaxies as given by Shen et al.\ (2003) because of the biases
in their data which are explained in Graham \& Worley (2008).  Lange et
al.\ (2015) have however fit the double power-law model from Shen et
al.\ (2003) to the galaxy size-mass data from several thousand nearby ($0.01
\le z \le 0.1$) early-type (morphologically-identified diskless) galaxies
taken from the Galaxy And Mass Assembly (GAMA) survey (Driver et al.\ 2011).
The galaxy magnitudes had been converted into stellar masses by Taylor et
al.\ (2011), and the effective half light galaxy sizes derived by Kelvin et
al.\ (2012) who fit single S\'ersic models to images in ten bands
($ugrizZYJHK_s$).  We have taken the $K_s$-band results from Lange et
al.\ (2015) and show their mass-size relation in Figure~\ref{Fig3}a.  Given
that Figure~6 from Lange et al.\ (2015) reveals that the double power-law
cannot fully capture the curvature in the GAMA data at $M_* >$ 1--2$\times
10^{11} M_{\odot}$ --- where galaxies have larger radii than the double
power-law model --- we have therefore additionally included the mass-size
relation from Graham et al.\ (2006, see Eq.2.11 in Graham 2013).  While
Figure~\ref{Fig3}a reveals that the mass-size relation of local early-type
{\it galaxies} have larger radii at a given mass than the distant spheroids,
it also reveals that the masses and sizes of our local {\it bulges} overlap
with those of the distant spheroids (Damjanov et al.\ 2011).  Our data appears
more clustered in Figure~\ref{Fig3}a simply because we have been stricter with
the mass and size limit for our local bulge sample. 


In Figure~\ref{Fig3}b we show the sizes and S\'ersic indices of our local
compact massive systems, and the overlap with 
galaxies in the redshift interval 1.4 $< z <$ 2.7, as given by Damjanov et
al.\ (2011, their Table~2). 
We have excluded from Damjanov et
al.\ those galaxies with no reported S\'ersic index and those fit 
with an $R^{1/n}$ surface brightness model having a fixed value of $n$, such
as 4.  
As with our bulge data, the high-$z$ galaxy data in Damjanov et 
al.\ has also been taken from a compilation of different surveys: 17 in
their case.  Their galaxy selection criteria is thus varied, and it
can be seen in Figure~\ref{Fig3} to contain a much larger range of sizes and
masses than our data. 
%
%
%
Comparing the data points, we conclude that not all compact, massive spheroids need to
have undergone significant structural and size evolution since $z \lesssim 2.5$.

\subsection{The number density of local, compact massive spheroids} 

Twenty one of the 22 systems (20 bulges plus 2 galaxies) are within 90 Mpc,
with the additional system at 103.6 Mpc.  Excluding this furthest bulge gives
a number density of $6.9\pm1.5 \times 10^{-6}$ Mpc$^{-3}$ for the sample of 21.
All of these systems have stellar masses in the range $0.7\times 10^{11} < M_* / M_{\odot}
< 1.4\times 10^{11}$. 
Ten of the 13 systems with $M_* \ge 10^{11} M_{\odot}$ are within 70 Mpc,
giving a similar number density\footnote{The usual dependence on the Hubble constant
  --- in this case $h^3_{73}$ --- has been omitted given that the spherical
  volumes used have not been exactly matched to any particular outer-most
  galaxy's distance.} of $7.0\pm2.2 \times 10^{-6}$ Mpc$^{-3}$.  
It needs to be remembered that these densities are a lower limit because we
have not conducted a volume-limited, all-sky survey for compact massive
spheroids.  However, the ATLAS$^{\rm 3D}$ survey did sample all of the bright
galaxies over half the sky to a depth of 42 Mpc.  Given that it contains three
compact, massive systems, this corresponds to a number density of
$(1.9\pm1.1)\times10^{-5}$ Mpc$^{-3}$. 
This value is 2.75($\pm58$\%) times higher, but there is a larger 
uncertainty assuming Poissonian errors.

While Taylor et al.\ (2010) reported a reduction, at fixed size and mass, of
at least 5000 in the co-moving number density of compact massive galaxies from
$z\sim 2.3$ to $z\sim 0.1$, we find that there is a roughly comparable (within a 
factor of a few) number density of 
compact massive systems at $z = 0$ and $z\sim 2.5$.  Bezanson et al.\ (2009)
report a number density of $(3\pm1)\times10^{-5}$ Mpc$^{-3}$ at this higher
redshift, for systems with stellar mass densities greater than one billion
solar masses within their innermost sphere of radius 1 kpc (or $M_* > 10^{11}
M_{\odot}$, $R_{\rm e} < 2.88$ kpc).  Muzzin et al.\ (2013, 
their Figure~5) report a number density of $\sim 5\times10^{-5}$ Mpc$^{-3}$ at
$2 < z < 3$ for all (compact and extended) quiescent galaxies with $M_* \sim
10^{11} M_{\odot}$, in fair agreement with the value of $\sim 4\times10^{-5}$
Mpc$^{-3}$ for the quiescent galaxies at $z=3\pm0.5$ with $M_* >
0.4\times10^{11}\, M_{\odot}$ reported by Straatman et al.\ (2014).
At these high redshifts, it is speculated here that the bulk of the disk formation
(which will remove ``galaxies'' from satisfying the compactness criteria), may
be yet to occur and thus one may have a cleaner sample of `naked-bulges' with
which to make a comparison with the number density of bulges in the local
universe.  At yet higher redshifts the `naked-bulges' may likely still be 
developing themselves (e.g.\ Dekel \& Burkert 2014). 

Barro et al.\ (2013) report a similar number density as Bezanson et
al.\ (2009) at $z=2.5$, even though Barro et al.\ used a lower stellar mass
limit of 0.1$\times10^{11} M_{\odot}$ for their sample (cf.\ 1.0$\times10^{11}
M_{\odot}$ used by Bezanson et al.\ 2009).  Barro et al.\ reported that their
number density increased as the redshift dropped, peaking at about
2.3$\times10^{-4}$ Mpc$^{-3}$ by $z=1.2$, before dropping to lower densities
as the redshift decreased further.  Taken together, this suggests that the
most massive, quiescent spheroids were in place first, with the less massive
spheroids appearing later at lower redshifts (possibly connecting with the 
luminous blue compact galaxies forming at $z<1.4$, Guzm\'an et al.\ 1997). 
The above peak density at $z=1.2$ is, at least in part,
larger than the value we have found because Barro et al.\ included galaxies
having a much broader range in stellar mass than we did.  In a separate study,
van der Wel et al.\ (2014) used a stellar mass limit $>0.5 \times10^{11}
M_{\odot}$ and found a peak density at $z\sim1.2$ of 1$\times10^{-4}$
Mpc$^{-3}$.  

The mass range sampled in our study only spans a factor of two, from 0.7 to
1.4$\times10^{11}\, M_{\odot}$.  Our number density per unit dex in mass, as
recorded in mass functions, is therefore five times higher, giving
3.5$\times10^{-5}$ Mpc$^{-3}$ dex$^{-1}$ (or $\approx 10^{-4}$ Mpc$^{-3}$
dex$^{-1}$ if using the volume-limited ATLAS$^{\rm 3D}$ results) at $M_*
\approx 10^{11} M_{\odot}$.  These values are roughly 2--6 times lower than
that from Barro et al.\ (2013), whose galaxy mass range exceeded 1 dex, and
roughly 1--3 times lower than that from van der Wel et al.\ (2014), whose mass
range was less than 1 dex.  A proper comparison is however complicated because the
local bulge mass function is not quite flat from 0.1 to 1.4$\times10^{11}
M_{\odot}$, but increases as one moves to the lower-mass end (e.g.\ Driver et
al.\ 2007).  For this reason, our number 
density for local compact massive bulges, as estimated above, is expected to be less than 
the actual number density in the decade-wide mass range from say 0.14 to 
1.4$\times10^{11} M_{\odot}$.  Using a constant stellar $M/L$ ratio across
this mass range, to convert the bulge luminosity 
function in Driver et al.\ (2007) into a mass function, the actual number
density in this decade range might be $\sim$40\% higher, although predicting
this properly requires knowledge of the slope of the mass function for local {\it compact}
bulges.  We hope to perform a more complete, volume-limited investigation of
compact massive bulges within 100 Mpc in a forthcoming paper, enabling us to
construct the mass function of local compact bulges with $M_* > 10^{10}
M_{\odot}$.  This can then be better compared with the mass function at higher
redshifts.

The WIde-field Nearby Galaxy-cluster Survey (WINGS) has reported the existence
of many compact massive galaxies in nearby ($0.04 < z < 0.07$) clusters.
Including substantially lower mass galaxies than us, Valentinuzzi et
al.\ (2010) reported that 22\% of the WINGS galaxies with $0.3\times 10^{11} <
M_*/M_{\odot} < 4\times 10^{11}$ are compact, with a median size of
$1.61\pm0.29$ kpc (and a median S\'ersic index of $3\pm0.6$). For their
cluster galaxy sample they derived a lower limit (because they excluded field
galaxies) for the number density of $(1.31\pm0.09)\times10^{-5}$ Mpc$^{-3}$
within the co-moving volume between $z = 0.04$ and $z = 0.07$.  This density drops to
$(0.46\pm0.05)\times10^{-5}$ Mpc$^{-3}$ when they increase their lower mass
limit from 0.3$\times 10^{11} M_{\odot}$ to 0.8 $\times 10^{11} M_{\odot}$
(their Table~1).  They observed that the bulk of their compact galaxies are
lenticular galaxies, which are known to have $\sim$30\% smaller disk sizes in
galaxy cluster environments (e.g.\ Guti\'errez et al.\ 2004; Head et al.\ 2014).


Using the Padova Millennium Galaxy and Group Catalogue (PM2GC) spanning $0.03
< z < 0.11$, Poggianti et al.\ (2013a,b) explored outside of clusters --- 
most galaxies reside in the field or group environment --- 
and concluded that 
4.4\% of local field and group galaxies with stellar masses $0.3\times 10^{11}
< M_*/M_{\odot} < 4\times 10^{11}$ are compact, and that this
4.4\% --- most of which are lenticular galaxies --- has a median mass-weighted age
of 9.2 Gyr, and corresponds to a number density of $(4.3)\times10^{-4}$
Mpc$^{-3}$, or basically $3.2\times10^{-4}$ Mpc$^{-3}$ dex$^{-1}$ given their
mass range sampled.

Although the above two studies treated the lenticular galaxies as single
component systems when measuring both their ages and their circularized
half-light radii, we suspect that it is the bulges of these galaxies which are
the primary, compact massive spheroidal component.  In a sense, we may
therefore be dealing with the same type of galaxy as they are. 
While we do not try and 
resolve the question as to why their compact, massive {\it galaxies} were not found
by others who searched in the {\t SDSS} database, we do provide a couple of 
comments.  Disk galaxies viewed edge-on will of course have circularized
half-light radii which are notably smaller than the values obtained if they
are viewed face-on.  It may be of interest to see if the Valentinuzzi et
al.\ (2010) and Poggianti et al.\ (2013a,b) detections are dominated by
massive, edge-on disk galaxies.  It may additionally be interesting to see the
results (sizes, masses, number densities) after careful bulge/disk
decompositions have been performed for these WINGS and PM2GC samples. 

To recap our findings, we measure a lower limit of 
3.5$\times10^{-5}$ Mpc$^{-3}$ dex$^{-1}$ (or $\approx 10^{-4}$ Mpc$^{-3}$
dex$^{-1}$ if using the volume-limited ATLAS$^{\rm 3D}$ results) for compact
($R_{\rm e} \lesssim 2$ kpc) bulges with $M_* \approx 10^{11} M_{\odot}$.


\section {Discussion}

\subsection{The rise of disks}

As noted in the Introduction, within the literature, minor mergers are the
overwhelmingly preferred, albeit still problematic, solution to try and
transform the compact, massive spheroids at high redshifts into larger
spheroids by today.  While our discovery of numerous compact massive spheroids
at $z = 0$ suggests that this evolutionary path did not always transpire, we can
still ask about minor mergers.  If (wet or dry) minor mergers had built a
large fraction of each disk around today's compact massive spheroids, then it
may be telling us that there is something not quite right with $\Lambda$CDM
simulations\footnote{Combes et al.\ (2015a) provides an interesting review of
  simulations with MOND in regard to bulge formation.}.  This is because 
the simulations contain somewhat random
orientations of satellite galaxies which would not result in the formation of
a disk, although see Tempel et al.\ (2015). 
However if minor mergers have built the disks, the observations may then be
telling us that most central galaxies have had preferred `Great Planes' on
which their minor neighbors were located prior to infall and disk (rather
than bulge) building.  Some support for this idea can be found in the
disks-of-satellites around the Milky Way and Andromeda (Kroupa et al.\ 2005;
Metz et al.\ 2007; Ibata et al.\ 2013; 
Pawlowski et al.\ 2014).  What is of course different in our
scenario is that these Great Planes are not just a recent, local, phenomenon
but would need to have always been present over the past 10--13 Gyrs (see
Goerdt, Burkert \& Ceverino 2013). 

Hammer et al.\ (2005, 2009) have reasoned that many rotating stellar disks
could have been built in major gas-rich mergers, with the disk forming due to the net
angular momentum of the gas in the merger event 
(e.g.\ Barnes 2002; Robertson et al.\ 2006).  Even in gas-poor (dry) mergers,
Naab \& Burkert (2003) report that disks can form if the net angular momentum
is not canceled out (Fall 1979).  Given that 
rotating spiral galaxies at $z\approx 0.65$ are half as abundant as they 
are today (Neichel et al.\ 2008), disk growth since $z=2\pm0.6$ has obviously
occurred, and the outer regions of galaxies with prominant bulges are known to
be younger (e.g.\ P\'erez et al.\ 2013; Li et al.\ 2015). 
Arnold et al.\ (2011) additionally remarked that lenticular galaxies
might form through a two-phase inside-out assembly with the inner regions
built early via a violent major merger, and “wet” minor mergers (rather than
gas accretion) subsequently contributing to their outer parts. 

It is recognized that there has not been enough galaxy merger events to
transform the distant compact massive galaxies into today's {\it
  ordinary}-sized elliptical galaxies (e.g.\ Man et al.\ 2015), and therefore
the above mechanisms cannot, on their own, be invoked to fully explain the
growth of disks around the compact massive galaxies seen at high-$z$ such that
they have become today's lenticular galaxies with typical disk-to-bulge flux
ratios of 3 (i.e.\ bulge/total $=1/4$).  Observing higher gas fractions of
lower metallicity at higher redshifts, B\'ethermin et al.\ (2015) favour large
gas reservoirs over major mergers as the cause of the intense star formation
observed in massive galaxies at high redshifts.  Following the idea of
cyclical galaxy metamorphosis (White \& Rees 1978; White \& Frenk 1991; 
Navarro \& Benz 1991; Steinmetz \& Navarro
2002; see also Bournard \& Combes 2002), 
Graham (2013) advocated earlier suggestions that `cold' gas flows may
have contributed to the development of these disks, either via spherical
accretion of $\sim$10$^4$--$10^5$ K gas (e.g.\ Birnboim \& Dekel 2003;
Birnboim et al.\ 2007) or streams (e.g.\ Kere{\v s} et al.\ 2005; Dekel et
al.\ 2009; Kere{\v s} \& Hernquist 2009; Danovich et al.\ 2014).

This alternative process, which can operate in parallel with mergers (e.g.\ Welker
et al.\ 2015), involves 
the accretion of gas from not just the halo (e.g.\ Kauffmann et al.\ 1993,
their section~2.7; 1999, their section~4.5; Genzel et al.\ 2006), and 
possible molecular gas reservoirs for some galaxies (e.g.\ Tacconi et
al.\ 2008; Santini et al.\ 2014; Chapman et al.\ 2015), but also cold filamentary flows from the 
cosmic web (e.g.\ Ceverino et al.\ 2010, 2012; Goerdt et al.\ 2012; Rubin et
al.\ 2012; Stewart et al.\ 2013).  Although such cold gas accretion is yet to be
convincingly demonstrated as a {\it common} phenomenon, it has the power to
subsequently build a disk that transforms into stars.  
Indeed, it has been noted for well over a decade that the mass of a 
galaxy could double, due to gas accretion, in just a few Gyr (e.g.\ Katz et
al.\ 1996; Bournaud \& Combes 2002).  The discovery of very
low metallicity gas (0.02 solar) 37 kpc from a sub-$L_*$ galaxy with solar
metallicity by Ribaudo et al.\ (2011) revealed the presence of such gas, and
we now know that there is lots of low metallicity gas in the halos of galaxies
(e.g.\ Lehner et al.\ 2013; Prochaska et al.\ 2014).  Furthermore, Bouch{\'e}
et al.\ (2013) have revealed the inflow (as opposed to just presence) of this
material around a $z=2.3$ galaxy that has kinematics, metallicity and star
formation typical of a rotationally supported disk, and it is such that the
gas accretion rate is well-matched to the star formation rate (see also
Conselice et al.\ 2013).  It has also been established that there is a high
spatial covering fraction of Lyman-$\alpha$ gas clouds within 300 kpc of all
galaxy types at high-$z$ (Wakker \& Savage 2009; Prochaska et al.\ 2011; Thom
et al.\ 2011; Stocke et al.\ 2013; Tumlinson et al.\ 2013), and a review of 
cool gas in high~$z$ galaxies can be found in Carilli \& Walter (2013).
In particular, even quiescent elliptical galaxies can have a
massive cool reservoir around them (Thom et al.\ 2012; Tumlinson et al.\ 2013; 
Zhu et al.\ 2014; O'Sullivan et al.\ 2015). 
Although not surrounding a pre-existing compact, massive bulge, 
Prescott et al.\ (2015) report on the rotation of an 80 kpc 
Lyman-$\alpha$ nebula at $z\sim 1.67$ with an implied total mass of 
$3\times10^{11} M_{\odot}$ within a 20 kpc radius. 
While this is a developing field, extensive evidence for gas 
accretion at many redshifts is reviewed by Combes (2014, 2015b).

It seems likely that the compact massive galaxies would act as natural
gravitational seeds around which filaments or streams of cold gas would flow
inward and build disks that form stars.  The existence of early-type galaxies
with dual, large-scale counter-rotating stellar disks (e.g.\ NGC~4550, Rubin
et al.\ 1992; NGC~3032, Young et al.\ 2008) supports the idea of disk growth
via the accretion of external gas clouds (e.g.\ Coccato et al.\ 2014, and
references therein), although it may also be due to minor mergers.  Presumably
feeding is usually from the same direction, given the low frequency of
significant (by mass) counter-rotation, and after settling to the mid-plane
the rotation is aligned.  Small counter-rotating stellar disks and
kinematically decoupled cores do however reveal that this is not always the
case, and gas accretion can form interesting features such as warps, gas-star
misalignments and polar rings (e.g.\ Briggs 1990; Jog \& Combes 2009; Davis et al.\ 2011;
Mapelli et al.\ 2015).  However rather than only random accretion
orientations, Pichon et al.\ (2011) explains how cold streams can build a disk
in a coherent planer manner (see also Danovich et al.\ 2012, 2014; Prieto et
al.\ 2013; Stewart et al.\ 2013; Cen 2014; Wang et al.\ 2014), which is 
of key importance.  A second key aspect is the rapid 
formation of some of these disks (e.g.\ Agertz et al.\ 2009; Brooks et al.\ 2009) which 
would naturally explain the older ages of many lenticular galaxy disks today. 
It also implies significant galaxy growth independent
of the merging of distinct entities such as dark matter halos. 
This process of gas accretion is still observed today in massive 
early-type galaxies (e.g.\ Davis et al.\ 2011).

Disk growth in early-type galaxies is an
on-going phenomenon
(e.g.\ Yi et al.\ 2005; Kaviraj et al.\ 2007; Fabricius et al.\ 2014), 
albeit at lower levels today as less gas is available 
(e.g.\ Combes et al.\ 2007; Sage et al.\ 2007; Huang et al.\ 2012; 
Catinella et al.\ 2013).  The molecular gas usually 
resides in kpc-scale, rotating disks (e.g.\ Inoue et al.\ 1996; Wiklind et
al.\ 1997; Okuda et al.\ 2005; Das et al.\ 2005), as does the HI gas
(e.g.\ Serra 2012, 2014, and references therein).  Young et al.\ (2008), for
example, detail the slow ongoing growth of the disk in NGC~4526, a galaxy whose bulge
has a half light radius equal to 1.43 kpc (Krajnovi\'c et al.\ 2013) and a 
stellar mass\footnote{Based on $M_{r'}=-20.4$ mag and using $M/L_{r'}=5$
  (Krajnovi\'c et al.\ 2013).} of $0.55\times10^{11}$ (which, along with other
bulges, was not massive enough to be included in our sample). 
Perseus~A is yet another
example which is also still accreting cold gas (Salome et al.\ 2006).

As noted in Section~\ref{Sec_data}, NGC~1277, NGC~1332, NGC~5493 and NGC~5845
are examples of local early-type galaxies where the disk does not dominate the 
light at large radii, as in proto-typical lenticular galaxies.  
These disks are an order of magnitude larger than nuclear
disks, and referred to as intermediate-sized disks.  Given that the amount of 
gas accretion can vary, it would be natural for such disks to exist. They are not a new
phenomenon (e.g.\ Scorza \& van den Bosch 1998, and references therein), 
but some readers may not be familiar with their existence. 
At $z \approx 2$, the lower mass spheroids, with less gravitational pull, and
which likely formed from a smaller over-density in the early universe, may naturally
experience a smaller subsequent supply of gas from cold streams and build
their disks more gradually.  Free-floating, ``compact elliptical'' dwarf
galaxies (e.g.\ Huxor et al.\ 2013; Paudel et al.\ 2014)
might represent spheroids which never acquired a significant disk, while 
those in the vicinity of much larger neighbours are thought to have been
largely stripped of their disks.

The theoretical work of Steinmetz \& Navarro (2002, and references therein) 
suggested that a galaxy's morphology is a transient phenomenon.  Galaxies do
not simply progress from the ``blue cloud'' to the ``red sequence'' (Faber et
al.\ 2007) but can move in the opposite direction.  While the latter pathway may not
be traversed in full today, because less gas is available, there are fledged examples of
galaxies in the `green valley' (e.g.\ Cortese \& Hughes 2009; Marino et
al.\ 2011).  Elliptical galaxies may initially be built through major mergers
of disk galaxies (and perhaps from a violent disk instability, Ceverino et
al.\ 2015), and then proceed to grow a new stellar disk through gas accretion
(White \& Frenk 1991, p.77), 
which remains intact until the next significant merger (see Salim et al.\ 2012
and Conselice et al.\ 2013 for supporting arguments, and Fang et al.\ 2012 and
Bresolin 2013 for caution in some cases).  The number of cycles may however be
low (i.e.\ 1 or 2) rather than several (3 to 5).  
Evidence for a bulge-then-disk scenario may exist within the Milky Way
(e.g.\ Zoccali et al.\ 2006).  Furthermore, the hierarchical models from
Khochfar \& Silk (2006), for example, present diskless galaxies at $z=2$ which
evolve into disk galaxies with a bulge-to-total mass ratio equal to 0.2 by
$z=0$ due to gas accretion from the halo, cold streams, and minor mergers.

There have recently been reports of infant, premature disks detected in some
of the high-redshift, compact massive galaxies (Chevance et al.\ 2012, and
references therein).  
%
%
In a study of 14 compact massive galaxies at $z=2.0\pm0.5$, van der Wel et
al.\ (2011) reported that 65($\pm$15)\% are disk-dominated, appearing highly
flattened on the sky and having disks with a median half-light radius of 2.5
kpc.  This corresponds to a median scale length of 1.5 kpc, which is half the
size of disks today, e.g.\ Graham \& Worley (2008, their Figure~3).  These
high-$z$ disks were observed to harbor compact central components,
i.e.\ bulges.  
Although most early-type galaxies around us today are known to be lenticular
disk galaxies --- thanks to studies such as ATLAS3D$^{\rm 3D}$ (Emsellem et
al.\ 2011) --- there are of course still some massive, slow or non-rotating,
elliptical galaxies.  Some of these may be the 'ordinary-sized' elliptical
galaxies\footnote{If so, one {\it may} need a mechanism to curtail disk
  growth, such as more efficient AGN feedback due to their lower stellar
  density, or hot X-ray halos.}, observed at high-$z$, while others likely
  formed from more recent, major merger events.  Some of the 'ordinary-sized'
  elliptical galaxies at high-$z$ may have also acquired disks, but they would
  appear as intermediate-sized disks if insufficient gas was accreted to build
  a larger scale disk around an already large galaxy. 

Following the deep observations by Szomoru et al. (2010) to verify the
compactness of a galaxy at $z=1.91$ found by Daddi et al.\ (2005), we note
that future work should be mindful that shallow surveys at any redshift could
miss the outer disks of galaxies to varying degrees and may largely just
recover the inner bulge if they are particularly shallow.  If that was to
occur, one would effectively, although accidentally, manage to identify higher
number densities of compact massive systems.


\subsection{Stellar Ages, and stellar mass loss} 

The above scenario for disk growth around compact ``bulges'' requires the
massive bulges of nearby disk galaxies to be old.  Although beyond the scope
of the current investigation, it will be of interest to explore the ages of
the bulges and disks in the current sample of galaxies.  However MacArthur,
Gonz\'alez \& Courteau (2009) have already shown that bulges in both early-
and late-type disk galaxies do indeed have old mass-weighted ages, with less
than 25 per cent by mass of the stars being young, second or third generation
stars built from metal enriched gas.  Based on stellar populations and radial
gradients, MacArthur et al.\ (see also Fisher et al.\ 1996) concluded that
early-formation processes are common to bulges and that secular processes or
`rejuvenated' star formation generally contributes minimally to the stellar
mass budget of bulges (see also Moorthy \& Holtzman 2006; Thomas \& Davies
2006; and Jablonka et al.\ 2007), yet it has biased luminosity-weighted age
estimates in the past.  Such 'frostings' of young stars, of up to 25 per
cent by mass, can give the impression of positive luminosity-weighted age gradients,
i.e.\ bulges are younger at their centres, and have misled some studies (which
assumed a single stellar population) into missing the fact that the bulk of
the stellar mass in most bulges is old\footnote{A caveat is that disks are
  built both around {\it and} within `bulges'.  The surface mass density of
  disks is higher in their centers than their outskirts. Therefore, disks
  which are relatively younger than their bulges will contribute younger stars
  where the bulge resides, but the dominant stellar population will be old at
  the center of the system.  It is of course also known that when disks do
  form, they may generate bars which can in turn further build on
  the``bulge'' via secular processes.  At the same time, strong bars can
  stall the inward accretion of gas, restricting it to the outer parts of the disk
  until the bar weakens (Bournaud \& Combes 2002).}.  
This follows Kuntschner \& Davies (1998) who revealed that the obvious
lenticular galaxies in the Fornax cluster have ages which are younger than the
more spheroid-dominated galaxies in the cluster.   


These works imply that the bulk of the stellar mass in today's massive bulges
already existed at high redshifts, and therefore these stellar systems should
be what we are observing in deep images of our young Universe.  Curiously,
Saracco et al.\ (2009) reported that there are two kinds of early-type galaxy
at $z \approx 1.5$: an old ($\sim$3.5 Gyr) population which needs to
experience a factor of 2.5--3 size evolution to have sizes equal to today's
early-type galaxies, and a young ($\sim$1 Gyr) population ({\it perhaps} those
which have recently acquired their disks) which already have sizes consistent
with today's population of early-type galaxies.  In this regard, it will be
interesting to know if their old population is best described with a single
S\'ersic model, while their young population is better described by a
bulge+disk model rather than a single-component S\'ersic model.  Furthermore,
if a younger disk has formed, or bulges developed at different epochs, they
may have different initial stellar mass functions (IMFs), resulting in
differing composite IMFs for the early-type galaxies today (e.g.\ Dutton et
al.\ 2013, McDermid 2015, and references therein).




Gradual stellar mass loss due to stellar winds is a part of the ageing process
for passively evolving spheroids (e.g.\ de Jager et al.\ 1988; 
Ciotti et al.\ 1991; Jungwiert et al.\ 2001), as is the conversion of 
visible stars into dark remnants such as neutron stars and stellar mass black
holes.  If a fraction $x$ of the initial mass is lost from a galactic system
due to stellar winds, or Type Ia supernovae clearing out gas, then there will
be an adiabatic expansion because the galaxy is no longer as tightly bound and
it will therefore reach a new equilibrium such that its size has increased by
$1/(1-x)$ (e.g.\ Jeans 1961; Hills 1980, their eq.6).  Of course if any
stellar ejecta remains in a galaxy --- which seems likely in massive galaxies
with strong potential wells ---, perhaps eventually ending up as hot X-ray
gas, then the expansion of the galaxy will be reduced depending on the radial
expanse of this gas.  After the young stars ($< 1$ Gyr) have evolved, stellar
winds likely account for just a $\lesssim$10\% reduction to the stellar mass
of the passive, compact massive galaxies seen at $z\sim2$ (Damjanov et
al.\ 2009, but see Poggianti et al.\ 2013b).  Fan et al.\ (2008) had
previously suggested that AGN feedback may blow out the gas in distant massive
galaxies and cause them to expand by factors of 3 or more.  Of course if this
had happened, then we would be left having to explain where all of the compact
massive bulges in the Universe today came from.
%

\subsection{Stellar density}

Given the smaller half-light radii that early-type {\it galaxies} had at
higher redshifts, before they acquired their disks, the stellar densities
within those half-light radii were 1 to 2 orders of magnitude greater than the
stellar densities inside the half-light radii of today's large early-type
galaxies.  This led Zirm et al.\ (2007) to conclude that it is a problem for
models of early-type galaxy formation and evolution. 
%
%
Buitrago et al.\ (2008) claimed that within the inner 1 kpc of the high-$z$
galaxies, they have densities equal to globular clusters.  This led them to
advocate a scenario in which globular clusters and distant compact massive
galaxies may have a similar origin, while at the same time suggesting that the
more massive halos (not the globular clusters) started collapsing earlier and
dragged along a larger amount of baryonic matter that later formed stars.
However they overestimated the typical globular cluster size by a factor of
10/3, and thus under-estimated a typical globular cluster's density by a
factor of 37, undermining their conclusions.

Bezanson et al.\ (2009) showed that the inner regions ($< 1$ kpc) of today's
early-type galaxies have stellar densities which are 2--3 times lower than
those of the distant compact massive galaxies of the same mass.  However, the
majority of galaxies in our sample have bulge-to-total ratios of 1/2--1/4
(Table~\ref{Tab2}), while Laurikainen et al.\ (2010) found an average ratio of
1/4 for their sample of lenticular galaxies.  That is, the local massive bulges are
2--4 times less massive than the galaxy within which they reside.  In
comparing {\it galaxies} of the same mass in Figure~2d from Bezanson et
al.\ (2009), one is effectively 
comparing the high redshift galaxies with local {\it bulges} that are 2--4
times less massive, possibly accounting for the lower density 
observed in these lower-mass bulges.  If one compares the quiescent, compact
massive galaxies at high-redshift with today's bulges of the same mass, one 
may find that they have the same density within the inner kpc. 

Hopkins et al.\ (2009) reported that there is no central ($< 1$ kpc) density
mismatch (see also van Dokkum et al.\ 2014). 
They suggested that ``the entire population of compact, high-redshift red
galaxies may be the progenitors of the high-density cores of present-day
ellipticals'', by which they mean spheroids.  We however consider the
evolution to have been very different, such that a 2D disk is built within and
around the spheroid that undergoes no substantial size-evolution, rather than the
continual development of a 3D envelope around a spheroid that effectively
undergoes significant size growth.

\subsection{Velocity Dispersions}

If there is indeed no evolution of the distant spheroids, then their velocity
dispersion should remain the same.  Now, if a stellar disk builds within and
around them, then the {\it galaxy} mass today will have increased.  As a
result, if one was to compare the velocity dispersions of the distant
compact spheroids with the velocity dispersions of $z = 0$ disk galaxies having the same
total stellar mass, one would actually be sampling local bulges which are less 
massive than the distant spheroids.  The local galaxies
of the same mass, containing compact bulges of 2--4$\times$ lower mass, 
are naturally expected to have {\it lower} velocity dispersions.  Indeed, this
general behavior has been observed (e.g.\ Cenarro \& Trujillo 2009; Cappellari et 
al.\ 2009; Newman et al.\ 2010). 

If one assumes $M_* \propto \sigma^4$, then a 2--4$\times$ difference in
stellar mass will be associated with a $\sim$1.2--1.4$\times$ difference in
velocity dispersion.  If one attempts to bypass the photometric stellar mass
estimate and use a dynamical / virial mass estimate ($\sigma^2 R_{\rm e,gal}$)
for the lenticular galaxies, the same issue occurs.  This is because of the
enhanced $R_{\rm e,gal}$ value due to the presence of the disk component.  If
a galaxy's half-light radius is 3--6 times larger than the half-light radius of
its central bulge, then the velocity dispersion of the bulge coupled with the
galaxy size will produce a galaxy virial mass which is 3--6 times larger than
what one would obtain for the bulge.  This will act to artificially separate
the compact high-$z$ spheroids from local massive bulges in a diagram of
dynamical mass versus velocity dispersion.  A note to keep in mind is that
disk growth, as with envelope growth, may lead to the compression of the
original inner bulge and a slight enhancement in the central velocity
dispersion (Andredakis 1998; Debattista et al.\ 2013).

\subsection{Depleted cores}

The presence of partially depleted cores in massive spheroids is thought to be
a result of major, dry merger events (e.g.\ Begelman et al.\ 1980; Faber et
al.\ 1997; Merritt et al.\ 2007).  The large elliptical galaxies with sizes
several times that of the high`$z$, compact massive spheroids, and with a
deficit of stars over their inner tens of parsecs to a few hundred parsecs,
likely formed this way.  The orbital decay of the binary supermassive black
hole, created by the galaxy merger, proceeds by slingshotting the central
stars out of the core of the newly formed galaxy.  However, as noted by Dullo
\& Graham (2013), the presence of such partially depleted cores in the bulges
of massive disk galaxies (e.g.\ Dullo \& Graham 2013) presents a conundrum
because these disk galaxies may be unlikely to have formed from such major
merger events\footnote{Major galaxy collisions may be more inclined to result
  in elliptical galaxies than disk galaxies, although see Naab \& Burkert
  (2003).}.  One potential solution is that the bulges may have formed first,
and the disks were subsequently accreted and grew (Graham 2013; Driver et
al.\ 2013).  It may therefore be interesting to check for partially depleted
cores in the current sample of disk galaxies in Table~\ref{Tab1}.  Application
of the core-S\'ersic model (Graham et al.\ 2003) to {\it Hubble Space
  Telescope} images --- due to their superior spatial resolution over a
sufficiently wide field of view --- will reveal which bulges have an inner
deficit of flux relative to the inward extrapolation of their outer S\'ersic
light profile.  Knowledge of which bulges have partially depleted cores,
coupled with information about their stellar age, should shed even further
light on the formation history.

\subsection{Formation paths and a cautionary remark on the $n=2$ or $n=2.5$ division of bulges}

Surveying predominantly compact, passive, spheroidal galaxies at $1.4 < z <
2.0$, with $10^{10} < M_*/M_{\odot} < 10^{11}$, Cimatti et al.\ (2008, see 
van Dokkum et al.\ 2008 for higher redshifts and higher masses) 
noted that the galaxy sizes are typically such that $R_{\rm e} \lesssim 1$
kpc, and are thus much smaller than early-type galaxies of comparable mass in
the present-day Universe.  However, the bulges of most disk galaxies at
$z\approx 0$ have compact sizes with $R_{\rm e} \lesssim 2$ kpc and many
bulges with $10^{10} < M_*/M_{\odot} < 10^{11}$ have $R_{\rm e} \lesssim 1$
kpc (e.g.\ Graham \& Worley 2008; Graham 2013, his Figure~1).  The growth of
disks in and around the compact, high-$z$ spheroids could therefore explain
their apparent disappearance by today. 

This possible explanation may at first glance appear somewhat at odds with the
claim in the interesting paper by van Dokkum et al.\ (2013) that disk models, 
in which bulges were fully assembled at high redshift and disks
gradually formed around them, can be ruled out.  However the first distinction
in these two remarks is of course that disks do not simply form around
spheroids but are additionally embedded within them; which is why bulge/disk
decompositions continue the disk all the way into the centers of galaxies.  It
would therefore be of interest to perform a bulge/disk decomposition of the
average surface density profiles, at different redshift intervals, that were
constructed by van Dokkum et al.\ (2013, their Figure~3).  This would allow
one to check how well their $z=0$ (Milky Way)-like galaxy profile that they
built resembles 
the Milky Way, and thus to know if their stacked profiles are reliable or
perhaps effected by dust or some other issue\footnote{The lack of any central
  bulge within the inner kpc of the $z=0$ profile is at odds with (Milky
  Way)-like disk galaxies, and the high average S\'ersic index ($n=2.6\pm0.4$)
  also suggests that their average profile is not representative of (Milky
  Way)-like galaxies, but rather an intermediate-luminosity early-type galaxy
  which have a very 
  different bulge-to-total mass ratio.}.  A reliable decomposition of an
evolving population (free from `progenitor bias') would of course enable 
one to quantify how much the disk 
and bulge (including pseudobulge) components have changed between the
different redshifts.  The second point of distinction is that van Dokkum et
al.\ (2013) were referring to the inner 2 kpc radius of (Milky Way)-mass
spiral galaxies, rather than $10^{11} M_{\odot}$ mass galaxies.  The Milky Way
has a bulge mass of $0.91\times 10^{10} M_\odot$ and a bulge-to-total stellar
mass ratio of 0.15 (Licquia et al.\ 2014); it also has a half-light radius of
0.66 kpc (Graham \& Driver 2007).  Therefore, within an inner radius of 2 kpc,
the mass of (Milky Way)-like galaxies are dominated by the disk rather than
the bulge.  Due to the greater prominence of bulges in early-type disk
galaxies than in late-type disk galaxies (see Figure~21 in Graham 2001), the
inner regions of today's massive galaxies --- which host the once distant,
compact massive spheroids --- are expected to have changed less over time, as
already observed by van Dokkum et al.\ (2010, their Figure~6).

Within the bulge-then-disk growth scenario, the existence of the distant
massive spheroids having $n<2$ implies that local galaxies with bulges having
$n<2$ need not be pseudobulges formed from the secular evolution of a 
stellar disk.  This topic is reviewed in Graham (2013, 2014) where it is noted
that bulges with S\'ersic $n<2$ need not be pseudobulges even if they are also 
rotating (e.g.\ Eliche-Moral et al.\ 2011; Scannapieco et al.\ 2011; Keselman
\& Nusser 2012; Saha et al.\ 2012; dos Anjos \& da Silva 2013; Graham 2014;
Querejeta et al.\ 2015). 

In passing we remark that formation paths in galaxy clusters in which the
disks of spiral galaxies fade and lose their pattern, and possibly end up with
relatively brighter bulges (e.g.\ Johnston et al.\ 2012, and references
therein), to become lenticular galaxies can still operate. That is, lenticular
galaxies likely have more than one evolutionary path, as can their bulges
(e.g.\ Cole et al.\ 2000; Springel et al.\ 2005), and the presence of both 
classical and pseudobulges in the same lenticular galaxy would seem to support
this (e.g.\ Erwin et al.\ 2003).  At the same time, the preservation,
i.e.\ the lack of evolution, of compact massive spheroids from $z=2\pm0.6$
until today has implications for studies of the evolution of the (black
hole)-bulge mass scaling relations (reviewed in Graham 2015, and references
therein).  The $M_{\rm bh} / M_{\rm bulge}$ ratio would not be expected to
decrease in these particular systems since $approx 1.4$.

McLure et al.\ (2013) reported that $44\pm12$\% of their $z=1.4\pm0.1$ sample
of relatively passive galaxies (specific star formation rate, sSFR $\le 0.1$
Gyr$^{-1}$) with $M_* \ge 0.6\times 10^{11} M_{\odot}$ have a disk-like
morphology.  While at face value this claim may appear to support the notion
that flattened 2D disks have developed, their morphological (disk-like) claim
was based on a S\'ersic index divide at $n=2.5$ such that they labeled
galaxies with $n<2.5$ to be late-type galaxies.  They adopted this divide
because Shen et al.\ (2003) had used it to separate the bright, local galaxy
population. Shen et al.\ (2003) could do this because their {\it SDSS} sample
contained relatively few dwarf early-type galaxies which have $n<2.5$.
However bulges can have $n<2.5$.  Therefore, some/many(?) of the galaxies in the
sample of McLure et al.\ (2013) with $n<2.5$ may still be rather ``naked
bulges'', rather than disk-dominated galaxies.  Indeed, the similar
distributions for their passive sample with $n<2.5$ and $n\ge2.5$ in the
size-mass diagram (their Figure~8) would suggest this.    

The practice of considering all high-$z$, compact, massive objects to be disk
galaxies if their S\'ersic index is less than 2--2.5 is unfortunately rather
common (e.g.\ Hathi et al.\ 2008; Fathi et al.\ 2012; Chevance et al.\ 2012;
Muzzin et al.\ 2012) and may have effectively mis-led many researchers.  For
this reason we have provided this cautionary text against this practice.


\section{Conclusions}

The compact massive galaxies at redshifts $z \approx 2\pm0.6$ have similar
stellar masses and sizes (and thus stellar densities), and radial
concentrations of light, as bright bulges in local disk galaxies.  Moreover,
the number density is not different by hundreds or thousands but is within a
factor of a few.  We have identified 21 compact ($R_{\rm e} \lesssim 2$ kpc),
massive ($0.7\times 10^{11} < M_* / M_{\odot} < 1.4\times 10^{11}$) spheroids
within 90 Mpc, giving a number density of $6.9\times 10^{-6}$ Mpc$^{-3}$ (or
3.5$\times10^{-5}$ Mpc$^{-3}$ per unit dex in stellar mass).  This is
however a lower limit because we have not performed a volume-limited search.
Based on a sub-sample taken from a smaller volume-limited sample, this density
may be 2.75($\pm58$\%) times higher at around $\times10^{-4}$ Mpc$^{-3}$
dex$^{-1}$.

This observation eliminates the need to grow all of the high-$z$ compact
massive spheroids by a factor of 3 to 6 by $z=0$, a challenge which had
remained unexplained for the past decade.  Rather, many of these high-$z$
spheroids need to remain largely unchanged in order to match the massive
bulges in today's early-type disk galaxies.  Therefore, any study which may
have claimed to have accounted for the size growth of quiescent galaxies would
have inadvertently, and most likely unknowingly, introduced an equally
puzzling problem.  If all the distant, compact massive spheroids had evolved,
we would then be faced with a second unexplained mystery, namely, where did
the compact massive bulges in the local universe come from and why are they
not observed in deep images of the $z=2.0\pm0.6$ universe.

We conclude that stellar disks have since grown around many compact massive
spheroids observed at $z=2.0\pm0.6$ and in so doing transformed their
morphological type from elliptical to early-type disk galaxy (i.e.\ lenticular
galaxies and early-type spiral galaxies).  Following Graham (2013), we again
speculate that some of the less massive compact spheroids at high-redshifts
may now reside at the centres of late-type spiral galaxies, and/or are today's
compact dwarf elliptical galaxies - which were either largely stripped of
their disk or never acquired one.


\acknowledgments
The authors thank Francoise Combes, Luca Cortese, Karl Glazebrook and Barbara Catinella for
kindly reading and providing helpful comments on this manuscript. 
A.G.\ thanks the organizers and participants of the 
enlightening conference ``The Role of Hydrogen in the Evolution of Galaxies''
15-19 September 2014, Kuching, Malaysia (Borneo).  
This research was supported by the Australian Research Council through
funding grant FT110100263. 

This work is based on observations made with the IRAC instrument 
(Fazio et al.\ 2004) on board the Spitzer Space Telescope, 
which is operated by the Jet Propulsion Laboratory, 
California Institute of Technology under a contract with NASA.
This research made use of the NASA/IPAC Extragalactic Database 
(NED: http://ned.ipac.caltech.edu) 
and the HyperLeda database (http://leda.univ-lyon1.fr).

{}

\end{document}